\documentclass[conference]{IEEEtran}

\usepackage{graphicx}
\usepackage{epstopdf}%

\makeatletter
\def\ps@headings{%
\def\@oddhead{\mbox{}\scriptsize\rightmark \hfil \thepage}%
\def\@evenhead{\scriptsize\thepage \hfil \leftmark\mbox{}}%
\def\@oddfoot{}%
\def\@evenfoot{}}
\makeatother
\usepackage{amsmath}%
\usepackage{amsfonts}%
\usepackage{amssymb}%
\usepackage{graphicx}
\usepackage{wrapfig}
\usepackage{subfigure}
\usepackage{algorithm2e}
\usepackage[all]{xy}
\usepackage{sidecap}
\usepackage{url}

\newcommand{\qed}{\nobreak \ifvmode \relax \else
      \ifdim\lastskip<1.5em \hskip-\lastskip
      \hskip1.5em plus0em minus0.5em \fi \nobreak
      \vrule height0.75em width0.5em depth0.25em\fi}

\newcommand{\wini}{\mbox{$\mathsf{C}$}}
\newcommand{\lin}{\mbox{$\mathsf{S}$}}
\newcommand{\mal}{\mbox{$\mathsf{Mallory}$}}

\newcommand{\wndsize}{$\left|\textit{wnd}\right|$}
\newcommand{\wnd}{$\textit{wnd}$}
\newcommand{\cwnd}{$\textit{cwnd}$}

\newcommand{\scom}{\mbox{$\mathsf{www.s.com}$}}
\newcommand{\malcom}{\mbox{$\mathsf{www.mallory.com}$}}
\newcommand{\malcoma}{\mbox{$\mathsf{www1.mallory.com}$}}
\newcommand{\malcomn}{\mbox{$\mathsf{wwwn.mallory.com}$}}
\newcommand{\malcoman}{\mbox{$\mathsf{www1-n.mallory.com}$}}

\newcommand{\ignore}[1]{}
\newcommand{\fullppr}[1]{}

\newcommand{\wzs}{marios} 

\newcommand{\atkbw}{\alpha}
\newcommand{\ampfact}{\mu}

\begin{document}


\title{TCP Injections for Fun and Clogging}
\author{Yossi Gilad and Amir Herzberg \\ Department of Computer Science Bar Ilan University}

\maketitle

\begin{abstract}
We present a new type of clogging DoS attacks, with the {\em highest amplification factors achieved by off-path attackers}, using only {\em puppets}, i.e., sandboxed malware on victim machines. Specifically, we present off-path variants of the {\em Opt-ack, Ack-storm} and {\em Coremelt DoS attacks}, achieving results comparable to these achieved previously achieved by eavesdropping/MitM attackers and (unrestricted) malware. In contrast to previous off-path attacks, which attacked the client (machine) running the malware, our attacks address a very different goal: large-scale clogging DoS of a third party, or even of backbone connections. 

Our clogging attacks are based on off-path TCP injections. Indeed, as an additional contribution, we present {\em improved off-path TCP injection attacks}. Our new attacks significantly relax the requirements cf. to the known attacks; specifically, our injection attack requires only a Java script in browser sandbox (not `restricted malware'), does not depend on specific operating system properties, and is efficient even when client's port is determined using recommended algorithm. Our attacks are constructed modularly, allowing  reuse of modules for other scenarios and replacing modules as necessary. We present specific defenses, however, this work is further proof to the need to base security on sound foundations, using cryptography to provide security even against MitM attackers.  

\end{abstract}


\section{Introduction} \label{intro}

As the importance of the Internet grows, there is higher risk in disruption of its services, i.e., perform {\em Denial/degradation of Service (DoS)} attacks; already we see DoS attacks becoming a significant concern, with different motivations, including criminal, ideological (`hacktivism') and even as part of cyberwarfare.
We focus on DoS attacks which are based on {\em network clogging}, i.e., disrupting network services by causing excessive traffic; these attacks do not depend on vulnerabilities of the victim network or application. Network clogging DoS attacks are becoming larger and more common, and becoming a significant concern to corporations and even governments. Recent reports indicate attack rates in the order of 100 Gbps\cite{arbor:reports}; 60-86.5\% of the attacks targeted the infrastructure (layer 3), including the mitigation infrastructure itself\cite{survey:ProlexicReport11}, which is mostly based on providing overcapacity upon attack. 

All clogging attacks are based on sending information from many malicious agents; such agents are normally malware running on user machines, ranging from fully-privileged malware to limited-privileges malware (e.g., Android application) and to {\em puppet}, i.e., malware running within standard sandbox (such as script or applet). It is easier for attackers to collect {\em puppets} which are malicious scripts running in a browser's sandbox \cite{AATA08:puppetnets}, these only require users accessing the attacker website. 

In this paper, we show that even such mere puppets can be abused to launch devastating clogging DoS attacks. The attacks we present exploit behaviors of the Transmission Control Protocol (TCP), or more precisely, exploit puppets running TCP on different client machines, to cause huge amounts of traffic leading to congestion. TCP is the main transport protocol over the Internet, ensuring reliable communication between applications;
in particular, puppets are usually restricted to communication using standard TCP stacks, except possibly for communication to the site sending the malware. This makes it challenging to use puppets for clogging attacks, since TCP responds gracefully to congestion to ensure fair and efficient sharing of network resources. 

Trivially, a MitM attacker can eavesdrop, block, modify and inject fake TCP traffic; such attacker does not need any client agents to perform DoS (and other) attacks. However, we consider the weaker - and more common - {\em off-path attacker}, who can only send spoofed packets, and cannot block, intercept or modify traffic. We show that even such weak attacker, can exploit puppets to cause wide-scale congestion, with very high amplification factor (i.e., a relatively modest number of puppets suffice to clog large network links). 

Specifically, we show how puppets can be used to launch the two clogging-DoS attack which have the highest impact (amplification) so far, the {\em Opt-Ack} attack \cite{SBB05:OptAck} and the {\em Ack-storm} attack \cite{AH11:Ack}, but using only puppets and an off-path attacker (instead of requiring privileged malware and weak-eavesdropping abilities, as in the original attacks). Furthermore, we explain how such attacks on specific connections, can be combined to create effective congestion disrupting connectivity of whole regions, extending and improving upon the Coremelt attack \cite{conf/esorics/StuderP09}, again requiring only puppets and an off-path attacker, instead of the original requirement of privileged malware. 

The basic idea of our attacks is to use the off-path attacker to send control packets to the TCP connection between puppet and server, playing the role of the malicious client in Opt-Ack and of the eavesdropping attacker in the Ack-Storm attack.
To do this, our attacker must be able to inject traffic into the TCP connection; this is not trivial, since puppets are not given the necessary information about a connection, such as the client, port and sequence numbers (of the client and the server). We show, however, that the attacker can obtain all necessary information for the attack, namely, efficiently perform an {\em off-path TCP injection attack} into the connections between the puppets and the servers. 

Our attacks can target either end of the TCP connection (client or server). The injected data is of TCP control plane; hence, these attacks hold even if the SSL/TLS protection is employed. We show in experiments that using a relatively small set of low-privileged malicious agents, e.g., scripts in browser sandbox, the off path attacker can cause large amounts of traffic leading to congestion. 
 
\subsection{Off-path TCP Injection Attacks}
Over the years, TCP specifications and implementations were enhanced to provide security against off-path attacks.
TCP implementations should randomize the 32-bit sequence number \cite{rfc1948} and the 16-bit client port \cite{rfc6056}; for successful injection, the adversary must provide valid values to these fields. 
Since the adoption of randomized initial sequence numbers and until recently, TCP was widely believed to be immune to off-path (injection) attacks. One exception was the off-path attacks on TCP of \cite{watson2004slipping}, which disconnected BGP connections that use constant client ports; countermeasures make this attack inapplicable today \cite{rfc4987}. However, this attack was considered as reflecting a very specific vulnerability of BGP availability, and 
the widespread belief was that an `off-path' spoofing attacker, cannot {\em inject} traffic into a TCP connection, since it is infeasible for the attacker to guess correct client port and sequence number. 
This belief is even stated in RFCs and standards, e.g., in RFC 4953, discussing on TCP spoofing attacks (see \cite{rfc4953}, Section 2.2). Indeed, since its early days, most Internet traffic is carried over TCP - and is not cryptographically protected, in spite of warnings, e.g., by Morris  \cite{Morris85} and Bellovin \cite{Bellovin:security:problems:in:TCP,Bellovin:Look:Back:at:TCP:IP:Security}. 

The first `proof of concept' showing that TCP injection attacks may still be possible, even with randomly-chosen initial sequence numbers, was in \cite{lkm:phrack:07}. 
This was recently improved into 
efficient off-path TCP injection attacks \cite{snptcp,woottcp,CCS12:tcp}; we could use these techniques to achieve our off-path clogging attacks. 

However, these existing attacks are limited, specifically, both \cite{snptcp} and \cite{CCS12:tcp} require malware rather than puppets on the client machine (albeit with limited privileges), 
while \cite{woottcp} exploits the use of globally-incrementing allocation mechanisms for both ports and IP-IDs, as exists in the popular Windows operating systems. 

We present new variants of TCP injection attacks, with further improved efficiency and avoiding both malware and globally-incrementing IP-ID and ports assumptions (see comparison in Table \ref{Tbl:blocks}). 
Although Windows is very popular, avoiding the globally-incrementing requirements still significantly expands the base of clients which can be exploited for the clogging attack; furthermore, these specific weaknesses may be removed in future versions of Windows. 
Our new attacks can also be used to extend the XSS, CSRF and phishing attacks of  \cite{snptcp,woottcp,CCS12:tcp} to these additional scenarios (puppet rather than malware and avoiding the requirement of Windows-specific properties).

\begin{table*}
\centering
\begin{tabular}{ c || c | c | c |}
    \cline{2-4}
                                 & Identify Victim-Connection         & Learn Victim-Connection Sequence Numbers           & Shown Exploit\\ \hline \hline
                                     
        Lkm \cite{lkm:phrack:07} & $\begin{array}{c}\text{Probe for connection} \\ \text{(Client runs Windows, no firewall)}\end{array}$ & $\begin{array}{c}\text{IP-ID side channel,} \\ \text{both seq. \# obtained} \\ \text{(Client runs Windows)}\end{array}$& None\\ \hline

    Zhiyun et al. \cite{snptcp,CCS12:tcp}  & $\begin{array}{c} \text{Monitor Connections with netstat/procfs}\\ \text{(Malware)}\end{array}$& $\begin{array}{c} \text{Exploit seq \# filtering,} \\ \text{only server's seq. \# obtained} \\ \text{(Malware; in \cite{snptcp} also sequence-checking firewall)}\end{array}$& $\begin{array}{c} \text{XSS, CSRF} \\ \text{(Malware, No TLS/SSL)}\end{array}$ \\ \hline
    	
    Gilad et al. \cite{woottcp}  & $\begin{array}{c} \text{Establish connection, exploits} \\ \text{sequential port allocation} \\ \text{(Puppet, client runs Windows)}\end{array}$ & $\begin{array}{c}\text{IP-ID side channel,} \\ \text{both seq. \# obtained} \\ \text{(Puppet, client runs Windows)}\end{array}$& $\begin{array}{c} \text{XSS, CSRF, phishing} \\ \text{(Puppet, no TLS/SSL)}\end{array}$\\ \hline
        
    This work                    & $\begin{array}{c}\text{Establish connection,}\\ \text{Timing side-channel} \\ \text{(Puppet)}\end{array}$
    & $\begin{array}{c}\text{``Inject and Observe'',} \\ \text{both seq. \# obtained} \\ \text{(Puppet, no TLS/SSL)}\end{array}$                        & $\begin{array}{c}\text{Denial of Service} \\ \text{(Puppet)} \end{array}$ \\
    \hline
    
\end{tabular}
\caption{Off-Path TCP Injection Attacks, Building Blocks and their Requirements (Specified in Brackets).}
\label{Tbl:blocks}
\end{table*}


\subsection{Attacker and Network Model} \label{intro:model}

\begin{figure}
  \begin{center}
    \includegraphics[width=0.35\textwidth]{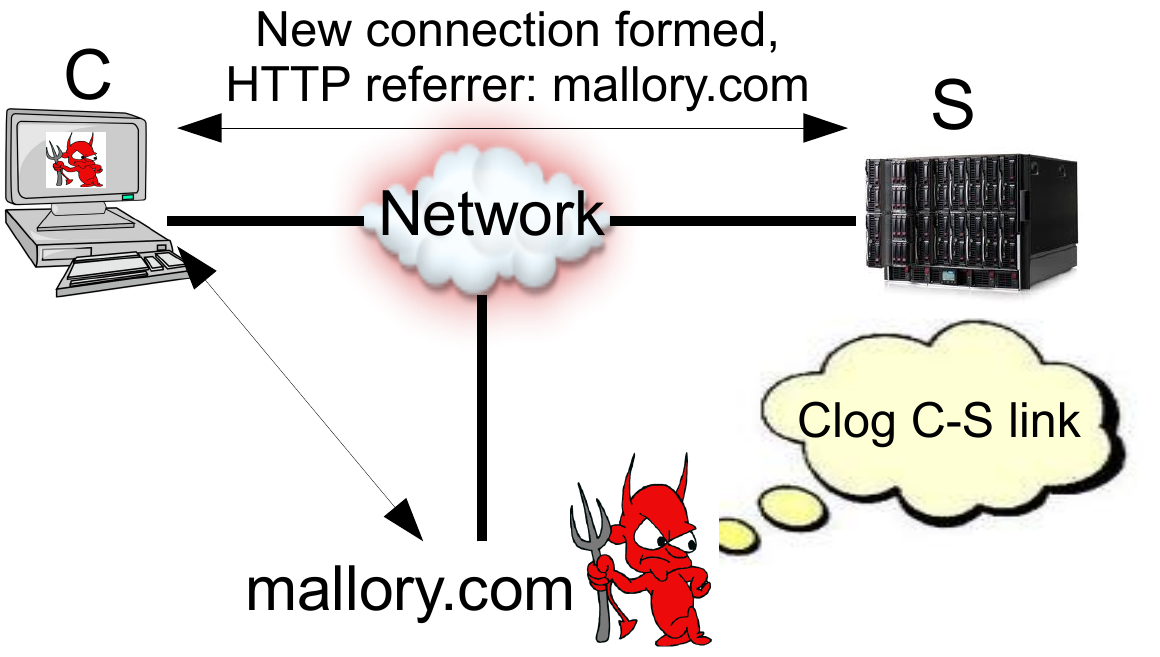}
  \end{center}
  \caption{Network Model. \wini\ enters \malcom, the adversarial web page. A script on that page forms a connection with \scom.}
   \label{fig:ourmodel}
\end{figure}

\mal, the attacker that we consider, is an off-path spoofing attacker. \mal\ cannot observe traffic sent to others; specifically, she cannot observe the traffic between a client \wini\ and a server \lin. However, \mal\ can send {\em spoofed packets}, i.e., packets with fake (spoofed) sender IP address. Due to ingress filtering \cite{rfc3013} and other anti-spoofing measures, IP spoofing is less commonly available than before, but still feasible, see \cite{SpooferProject, journals/toit/EhrenkranzL09}. Apparently, there is still a significant number of ISPs that do not perform ingress filtering for their clients (especially to multihomed customers). Furthermore, with the growing concern of cyberwarfare, some ISPs may intentionally support spoofing. Hence, it is still reasonable to assume spoofing ability. Spoofed packets were used in many other attacks, including SYN-flood, DNS-poisoning and both previous off-path injection attacks. Figure \ref{fig:ourmodel} illustrates our model.

\subsection{Contributions}
This first contribution of this paper, is in presenting the very effective off-path clogging DoS attacks, significantly more efficient than previously known off-path clogging attacks such as DNS reflection attacks. Furthermore, our attacks are the first to show that off-path clogging can utilize TCP traffic. 

Third, we investigate off-path variants to previous DoS attacks which were known to require stronger adversaries: either man-in-the-middle, or control of client end (i.e., bot). We present an algorithm that allows an off-path adversary to perform the Coremelt attack and disconnect a targeted autonomous system from the Internet. We show that this attack has significantly lower requirements (of the number and location of agents) from the original Coremelt attack.

A second contribution allows an off-path attacker to efficiently {\em detect the client source port}, if the client's operating system uses the {\em Simple Hash-Based Port Selection Algorithm} recommended in \cite{rfc6056}. This mechanism is deployed in Linux and is extensively used, e.g., by Android; moreover, many clients connect to the Internet via NATs, these devices commonly run Linux and use this selection algorithm for the external port, making it the de-facto port selection algorithm of many clients. Detection of the source port is a necessary first step for efficient TCP injection, but also has additional security implications, in particular, allowing traffic analysis (detection of which clients connect to a server), as investigated in \cite{SpyingInTheDark}. 

%

Our third contribution is a new, efficient off-path TCP injection attack, where an off-path adversary learns both the server and client's sequence numbers of the connection using only a puppet. This improves over \cite{snptcp,CCS12:tcp} that assume malware running on the victim machine (although with low privileges) and only exposes the server's sequence number. Our attack is also not specific to TCP implementation (platform) as the technique in \cite{woottcp}.

One last contribution is in introducing and modeling the problem of clogging DoS attack optimization, which provides a better measure of the potential risk of clogging due to coordinated attacks utilizing many agents. 



\subsection{Organization}
The reminder of this paper is organized as follows. Section \ref{background} presents an outline of our attacks, and `breaks' them to modular structure of building blocks. Section \ref{clientport} presents our port de-randomization algorithm. Sections \ref{serverseq} and  \ref{Injection:AckEsposure} presents our sequence number inference technique, for the server and client numbers, respectively. Section \ref{exploits:DoS} presents and evaluates the off-path variants of the Ack-Storm and Optimistic-Ack attacks and Section \ref{coremelt} builds on these attacks to perform a variant of the Coremelt attack with agents with low-privileged malicious agents. Section \ref{Injection:Defenses} presents defense mechanisms that can be deployed in either the client or the server. Finally, Section \ref{conclusions} presents our conclusions from this work.


\section{Background and Building Blocks} \label{background}

In this section we divide the TCP injection technique into building blocks which form a general scheme for performing an off-path TCP injection attack. This scheme is modular: each building block is a closed component, that can be implemented in several ways under different assumptions. Indeed previous off-path TCP injections \cite{woottcp,snptcp,lkm:phrack:07} follow this scheme. We present a high level description of each step and compare implementations of the same block in previous attacks. The reminder of this section describes each of the building blocks and compares their implementation in existing attacks. Table \ref{Tbl:blocks} presents a summary of the discussion below.

\subsection{Identify Victim-Connection} \label{establishconn}
Our first building block is to identify a TCP connection to attack, i.e., a victim-connection. In \cite{lkm:phrack:07} the adversary actively scans the client machine to existing connection with a particular server. As indicated within \cite{lkm:phrack:07}, the technique is often detected and blocked firewalls.
In \cite{snptcp}, \mal\ runs a malicious application on the client machine (Android). The adversary uses this application to monitor connections from the client machines (e.g., by executing netstat). 

In this work we assume that \mal\ runs a puppet, i.e., script in browser's sandbox and use this puppet to establish the connection to which we inject traffic. We refer to this connection as the `victim-connection'. A puppet can cause the browser to establish a connection with a server of choice by requesting an object such as an image to embed in a web page, e.g., \url{<}\text{img src}\url{="http://www.server.com/img.jpg"\>}. The same technique was used in \cite{woottcp}. 


Once the adversary identifies that a connection between the client and a server exists, she aims to learn the TCP four tuple, i.e., IP addresses and ports of both peers, these parameters identify the TCP connection. The adversary in \cite{lkm:phrack:07} receives the four tuple as result of probing. An adversary running malware as in \cite{snptcp} obtains this information directly from netstat output (or reading the corresponding file in procfs).

A puppet is more limited than malware and does not have access to operating system information. In both puppet based techniques (this paper and \cite{woottcp}) the adversary establishes the connection. Therefore, three of the parameters in the TCP four tuple are known: the client IP address, since the client is connected to the \malcom; the server IP address and port, since the attacker chooses the remote-end for the connection. Hence, it is only left to identify the client port. In \cite{woottcp} the authors exploit the sequential port allocation implemented in Windows; this predictable allocation mechanism allows the adversary to guess the correct port of the connection that the puppet established.

However, many operating systems make efforts to avoid predictable port allocation, as recommended in \cite{rfc6056}. Transport layer client port randomization is a common countermeasure against off-path (`blind') attacks; it was proposed as patch to many attacks, including TCP RST attack \cite{watson2004slipping} and DNS poisoning \cite{kaminsky:dns}. Since the adversary cannot observe the traffic, unpredictable client port provides an additional 16-bit randomness that the blind adversary must `hit' in order to tamper with the communication. 

We focus on one of the algorithms, {\em Simple Hash-Based Port Selection}, recommended in \cite{rfc6056} and implemented in Linux. This algorithm uses a random initial port for every destination; the port number is incremented for every connection. While \mal\ cannot observer the port that \wini\ uses to communicate with \lin\ (cannot eavesdrop), we show how she can use the puppet to: (1) open multiple connections between \wini\ and \lin, and (2) learn the value of one of the allocated ports. 

After such port is obtained, a binary search is performed to obtain the last (highest) port that the \wini\ allocated. This allows the adversary to identify client ports used in following connections between \wini\ and \lin\ since following client-ports will use the same counter.

Our technique is based on the Indirect Rate Reduction attack, that fakes TCP congestion events. This attack was suggested in \cite{SpyingInTheDark} to identify the existing of TCP connections via the Tor anonymity network.
We perform this rate reduction attack on different client ports and identify that a connection exists through a particular client port if the puppet observes a rate reduction, as follows from the TCP specification (cf. an operating system flaw as used in \cite{woottcp}). Our technique can possibly precede other off-path attacks to bypass port randomization protection (for TCP). 

\subsection{Learn Victim-Connection Sequence Numbers}

Once \mal\ has identified the victim-connection, she learns the sequence numbers; knowledge of these parameters allows her to inject data to the connection. 
Observing the sequence numbers directly from traffic requires either an on-path attacker or privileged malware. Therefore, this option is unavailable for the weak attackers considered in \cite{lkm:phrack:07,woottcp,snptcp} and this paper. Off-path TCP injection techniques provide different methods to infer on the sequence numbers. We briefly describe existing techniques and the one presented in this paper.

We present a new TCP server sequence number inference attack, that uses a radically different approach than the previous attacks. Specifically, the attacker sends to the client  data spoofed as coming from the server; by cleverly manipulating client queries, we are able to actually read such data when it falls into the flow control window. The data contains the server sequence number, hence, when read by our puppet, the puppet learns the server's sequence number. While this attack has to be done carefully to avoid connection reset, the result is extremely efficient attack. We next describe this technique, but before that, we recall for completeness the techniques used in previous works.

\subsubsection{Windows Specific, Bidirectional Exposure}

In the Windows TCP sequence number exposure attacks \cite{woottcp,lkm:phrack:07}, the adversary exploits the global counter IP-ID implementation in Windows. \mal\ observes the difference in the IP-ID field in packets that she receives from the client machine which tells the number of packets that the client machine sent to other destinations (since each packet increments the IP-ID).

In this technique, the attacker sends to the client crafted spoofed packets (appear to be sent from the server). The client responds to these packets only if they do specify incorrect server sequence number, i.e., outside the client's flow control window. The client sends the response to the server and \mal\ learns whether the client responded by observing the IP-ID field (in packets that she receives). After learning the server's sequence number, the techniques in \cite{woottcp,lkm:phrack:07} exploits Windows TCP implementation which filters incoming packets according to their acknowledgment numbers (this mechanism is not standard). This implementation allows the adversary to learn which acknowledgment is valid (passes filtering) using a similar side channel as the IP-ID. The acknowledgment number that the client expects to receive is close or equal to his sequence number.

\subsubsection{Unidirectional Sequence Number Inference Attack}
In the sequence number inference attack \cite{snptcp}, the adversary sends spoofed packets to the client machine. Each packet specifies a different, guessed sequence number. The observation in \cite{snptcp} is that if the sequence number is not close from the value that the client expects (i.e., the sequence number that server will next use), then a firewall connecting the client to the network might discard this packet. According to measurements in \cite{snptcp} 31.5\% of cellular carriers in the United-States deploy firewalls which filter TCP packets by their sequence numbers; hence, their costumers are vulnerable to the sequence inference attack.

The paper suggests two side channels that allow the adversary to learn the current value of the counter. The first aids globally incrementing IP-ID that might be deployed on routers and other intermediate middle boxes. The adversary sends a probe packet, spoofed as arriving from the server, that has TTL value high enough to pass the firewall, but low enough to be discarded by a following router; this packet will cause the router to send an ICMP feedback to the spoofed source. The adversary then sends a similar (but non-spoofed) packets to the victim and learns according to the difference in IP-ID whether the middle box responded to the probe.
Another information channel suggested by \cite{snptcp} is using the malware running on the client to directly read a system file (under procfs) that specifies the number of packets that arrived at the client machine since boot time. \mal\ learns that the sequence number specified in a probe passed firewall filtering if that counter increments (i.e., client receives probe). The malware monitors this variable and informs the attacker whether her packet arrived at the victim.

In this case that \mal's probe passes firewall filtering, she identifies that the sequence number specified in the probe is close to the correct sequence number. Once the attacker finds some sequence number in the firewall window, she efficiently finds the beginning of the window using a binary search. This attack exposes the server's sequence number, allowing the adversary to inject data as the server, but not the client.

\subsubsection{``Inject and Observe'', Bidirectional Sequence Number Inference Attack} \label{injectandobservehigh}

In this work we present a new technique that allows an off-path adversary to learn the value of both sequence numbers. Our technique assumes a standard TCP/IP implementation (c.f. to \cite{woottcp}) and does not rely on leakage via the IP-ID. Furthermore, since we assume only a puppet running on the client machine, the adversary cannot receive feedback from system files. 

Our technique is divided into two phases. In the first step, described in Section \ref{serverseq}, \mal\ learns the server's sequence number. 
In this step \mal\ tries to inject a probe packet to the victim-connection; if successful, the puppet receives the injected data which contains the sequence number specified in the probe. The probability that \mal\ exactly the next sequence number in the probe packet is low ($\frac{1}{2^{32}}$). However, \mal\ has a reasonable probability to specify a valid sequence number, one that is in the client's flow control window. We therefore send a probe that has meaningful parsing even buffered in the client's queue and is partially overrun by genuine segments from the server.

Successful learning of the server's sequence number is required to perform the second phase, described in Section \ref{Injection:AckEsposure}, where \mal\ learns the client's sequence number. 
We employ a similar approach, however, knowledge of the server sequence number allows sending packets that are discarded if they specify future acknowledgment number and accepted otherwise. This allows us to perform a binary search for the acknowledgment number that the client expects (which equals his sequence number).

In this step of the attack, the adversary sends a spoofed packet to the client as the server, that specified correct sequence number (learned in previous step) and an Ack number. This packet is discarded if its Ack number is higher than the client's sequence number (i.e., packet Acks something not yet sent) and reaches the application otherwise. This step can also extend the technique in \cite{snptcp} to allow bidirectional injection i.e., allow the attacker to inject data to the server-side as well.

\subsection{TCP Injection: Exploits} \label{dosandother}

TCP injections have several exploits. Zhiyun et al. \cite{snptcp} show that these can be exploited to perform cross-site scripting and request forgery, Gilad et al. \cite{woottcp} show an additional exploit for sophisticated phishing. In this paper we show a new type of exploit: enhancing denial of service (DoS) attacks. We show that by employing TCP injections, the adversary can perform DoS attacks using puppet-nets that are equally as powerful to botnets (i.e., controlled machines). As argued in \cite{AATA08:puppetnets}, it is relatively easy for attackers to control a large number of puppets (cf. bots).

Scripts running in browsers have limited ability to launch DDoS attacks. One reason is limitations placed by browsers on the number of concurrent connections opened by the browser. Another reason is that, due to their execution within a sandbox, these scrips can only use standard TCP connections. In particular, if the attack succeeds in causing congestion and loss, then TCP connections, including those of the puppets (scripts), will significantly reduce their window size (and hence their traffic rates). Therefore, while malicious Java Scripts can be used for DDoS attacks, their impact is much less than that of `regular' malware (zombies/bots). 
We show in Section \ref{exploits:DoS} how by using TCP injections, the adversary can persuade the server to send data in high rates, which is typically higher than the client's bandwidth.

In contrast to the data integrity exploits presented in \cite{snptcp} and \cite{woottcp}, the DoS exploits that we present use TCP control plane to cause congestion. Since there is no attempt to inject data to the application layer, even SSL/TLS protected sites are vulnerable. Hence, these exploits are applicable on websites that support HTTP connections, with or without SSL/TLS \footnote{For SSL/TLS websites, our exploit requires a different sequence numbers exposure technique, i.e., those in \cite{lkm:phrack:07,snptcp,woottcp}, as the one in this paper relays on injection of data.}.

\subsubsection{Denial of Service Exploits}
We show how a spoofing, off-path attacker, who controls a limited number of (weak) puppets, can deploy formidable DoS attacks, which so far were known to require  stronger attacker capabilities: the {\em Ack-Storm} attack \cite{AH11:Ack} and the {\em Optimistic-Ack} attack \cite{SBB05:OptAck}. Both are DDoS attacks, which use TCP control plane to generate excess amount of traffic. The Ack-Storm attack \cite{AH11:Ack}, is usually performed by MitM adversaries, possibly with limited eavesdropping abilities.  The Optimistic-Ack attack \cite{SBB05:OptAck} typically requires client cooperation (zombie) and persuades the server to send data in a high capacity, more than that allowed by \wini's link. 
Since in both these attacks \mal\ injects data only to the TCP layer (and not to the application), these attacks also work when the victim servers use SSL.


\subsubsection{Large Scale DoS Attack}
By launching the presented off-path clogging attacks cleverly, simultaneously on multiple client-server pairs, an attacker (\mal) to conduct an improved variant of the {\em Coremelt attack} \cite{conf/esorics/StuderP09} and congest a core link of the Internet, using only puppets. We explore this attack vector and introduce appropriate model and optimization problem for the attacker, to evaluate the most effective attack. 
Given a network graph, puppet locations and nodes to disconnect from one another, the attacker establishes victim-connections to servers, chosen according to their locations in the network. DoS attacks based on TCP injections are employed to disconnect victim nodes (e.g., an autonomous systems) from the network.

\section{Identify Victim Connection: A Puppet Only Technique} \label{clientport}

The first step in performing a TCP injection as described in the previous section is to identify the victim-connection. As described in Section \ref{establishconn}, since the adversary uses the puppet to establish the victim-connection, she knows the client address as well as the server address and port. In this section we describe a new technique that allows an off-path adversary running a puppet on a victim machine to learn the forth parameter of the TCP four tuple: the client port. 

As noted in Section \ref{establishconn} the client port is often randomized by the operating system. Larsen et al. suggest in RFC 6056 \cite{rfc6056} four client port allocation algorithms that are secure against a blind adversary trying to identify the client port. In this section we focus on the third suggestion: `Simple Hash-Based Port Selection'. This algorithm is used by the Linux kernel in versions 2.6.15 and above, i.e., from the year 2006; therefore, it is embedded in all Android versions. In Simple Hash-Based Port Selection, the operating system chooses a pseudo-random initial port for each destination (server), then for each new connection that the client establishes with that destination, the current port is used and then incremented. This protocol should be secure against off-path adversaries since these are not aware of the initial port. In the technique that we present below, the off-path adversary uses the puppet running on the victim to (1) establish multiple connections to the server, (2) time the server response time. 
The technique has two repeated steps which we describe below, iteration eliminates half the possible ports as used in the connections established by the puppet.

The technique that we present below is executed in iterations; each iteration eliminates half the possibilities for client ports used in the connections allocated by the puppet. An iteration is composed of two steps which: (1) connections establishment, and (2) port elimination.

\subsection{Step 1: Open Multiple Connections to \lin}
A script running in browser context does not open a connection directly; instead, the browser establishes a connection with the server \lin\ when the script requests an object from \lin\ (e.g., to dynamically embed a remote image to a web page). However, a new connection is established only if there is no existing connection with \lin\ (otherwise, request is sent on the existing connection). 

\begin{figure}
  \begin{center}
    \includegraphics[width=0.5\textwidth]{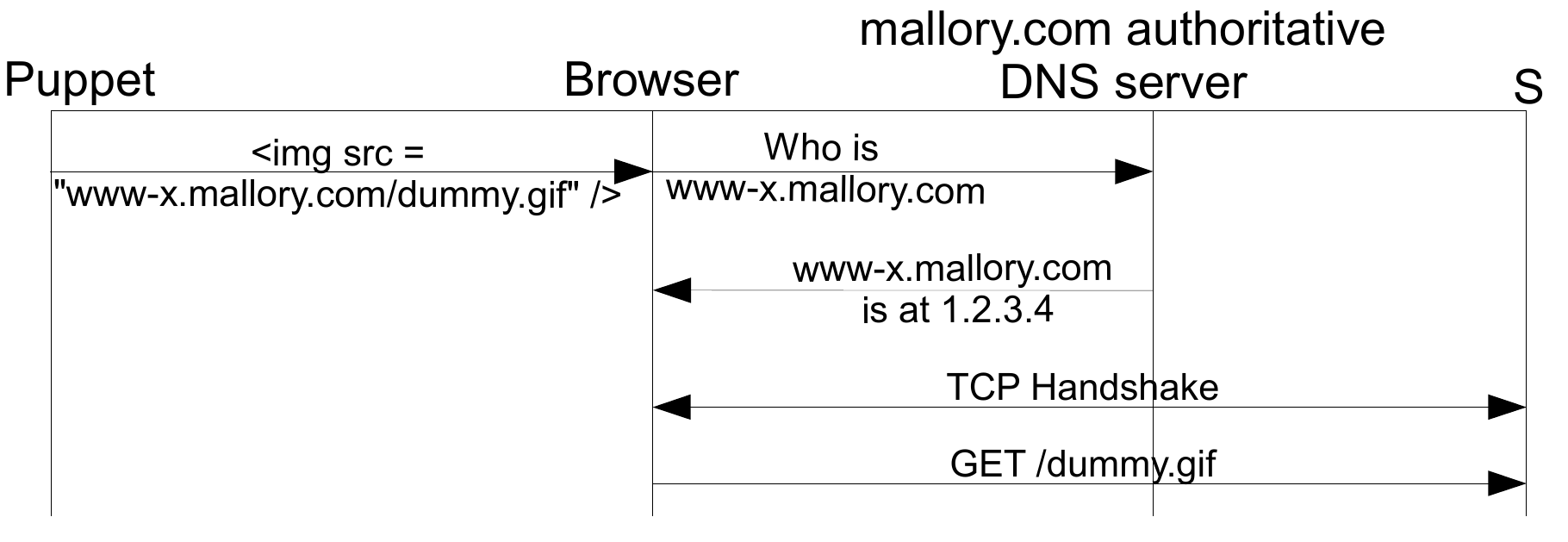}
  \end{center}
  \caption{Port De-randomization, Step 1: Establish new connection to \lin.}
    \label{fig:openconn}
\end{figure}

The adversary uses the puppet (script) to establish $n > 1$ connections to the server by manipulating DNS mapping of attacker controlled domains. The puppet requests an object from \malcoma,$\dots$,\malcomn; since \mal\ controls the DNS records for these domains, she sets the mapping of the domains to the same IP address, that of the \lin. Browsers use domain-names to identify servers (cf. to IP addresses); hence, this technique, which we verified on Firefox and Chrome, opens $n$ new connections to \lin, where $n$ is the maximal number of connections allowed (see value of $n$ in Subsection \ref{ports:values}). According to the port allocation algorithm, the ports used in these connections are sequential, we use this observation in the following step. Figure \ref{fig:openconn} describes this step of the attack.

Note that in the following iteration, when the puppet will open a new set of $n$ connections to \lin, connections opened in this step will be automatically closed by the browser, that usually closes the least recently used connection when out of resources (i.e., number of connections exceeds $n$).

\subsection{Step 2: Port Elimination}

\begin{figure}
  \begin{center}
    \includegraphics[width=0.5\textwidth]{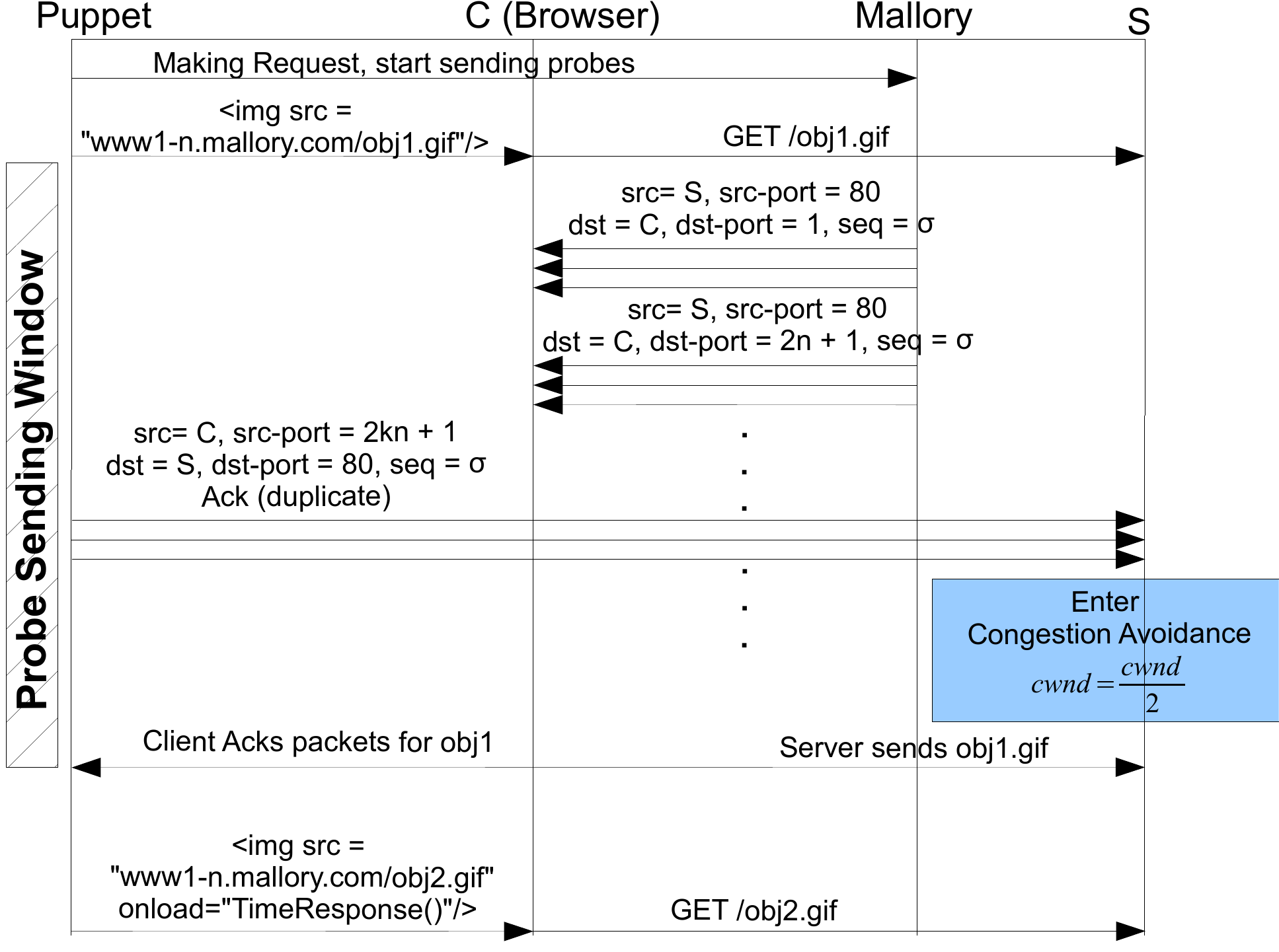}
  \end{center}
  \caption{Port De-randomization, Step 2: produce three duplicate Acks and measure rate; port $2kn + 1$ is a connection port.}
   \label{fig:portexposureiter}
\end{figure}

In this step, illustrated in Figure \ref{fig:portexposureiter}, \mal\ sends probes to the client \wini. These probes are spoofed and appear to be coming from \lin. The destination port (i.e., client port) specified in the probe packets is the one tested by the adversary. Probes are crafted to persuade the receiver to respond with a duplicate Ack if they belong a connection, i.e., the destination port is one of those allocated by the client machine in the first step. A sequence of three duplicate Acks that \wini\ sends \lin\ will slow the connection, this effect is measured by the puppet who notifies \mal.

A probe packet is a TCP packet that specifies a random sequence number; we assume the likely case, that this 32-bit field is out of \wini's flow control window. In this case, according to the TCP specification \cite{rfc793} (page 69), the receiver immediately responds to the source with a duplicate acknowledgment. A sequence of three duplicate acknowledgments is interpreted by TCP as loss and will cause the sender (\lin) to exit TCP slow start and enter congestion avoidance state. In TCP slow start, the congestion window (\cwnd), which decides the number of packets that can be transmitted in a round trip time (i.e., without receiving an Ack) grows exponentially, while in congestion avoidance it grows linearly. Furthermore, a loss event will significantly reduce the congestion window (up to half the original size).

In the elimination step \mal\ sends three probes to all ports $p = \textit{offset} \text{ (mod $2^in$)}$, where $i$ is the current iteration number and \textit{offset} is initialized to 1 and updated every iteration. In each iteration, the puppet requests a small object from all connection ports that she established, this by requesting an image from \malcoman. The puppet times each response and tests whether the time to load one object was significantly longer than others (in our experiments average measured increased by more than 120\%). The feedback that the puppet returns to \mal\ is a bit: $b = 0$ if there was a longer response, indicating that \mal\ had probed one of the connection ports, and $b = 1$ otherwise. 

The probes need to be synchronized with a puppet request for an object: although not written in the specification, we have found that many web-server (in particular those running Linux) validate when they receive three duplicate acknowledgments that there are at least three packets `in-flight' (i.e., un-acked) that they have sent to the client. Therefore, the restriction is that the client sends the duplicate acknowledgments after sending the request to the server. If there is no reordering in the network, this method ensures that the server will send the response before handling the duplicate acknowledgments. Notice that the client sends the duplicate Acks before actually receiving the server's response. We refer to this time frame in Figure \ref{fig:portexposureiter} as the \mal's probe sending window.

At the end of the step, \mal\ updates the attack state according to puppet's feedback: $\textit{offset} \leftarrow \textit{offset} + n + b \cdot 2^{i-1} n$. The reason that $n$ is always added to the offset is to account for the $n$ connections established in this iteration. After the final port elimination step, the value $p$ indicates a connection port.

\ignore{
Each iteration eliminates half the possible ports, after $\log_2(\textit{possible ports})$ the value $p$ is one of the ports that were allocated for the connection. This technique takes advantage on the algorithm sequential port allocation, allowing the adversary to only probe a fraction of the possible ports. Figure \ref{fig:portexposureiter} illustrates an iteration of the port exposure process.

The attacker can perform another $\log_2(n)$ iterations to identify the last allocated port and the current value of the port counter used for \lin. However, in the following phases of the TCP injection attack, we only require knowledge a client port used in a connection with \lin\ and not the exact counter value.
}

\subsection{Implementation and Limitations} \label{ports:values}

The Linux kernel chooses client port from the range $[32768, 61000]$ this is significantly a smaller range than all available ports, i.e., $[1,2^{16}-1]$. This observation to improve the search run time (only probing ports in the smaller range). The maximal number of connections that the puppet may open, i.e., $n$, changes according the version of the browser. However, this value is at least 32 in modern browser and typically increases with new releases \cite{browsers-stats}. These numbers imply that port de-randomization technique will require $\left\lceil  \log_2(61000 - 32768)\right\rceil = 15$ iterations to complete, and \mal\ will send $\left\lceil \sum_i^{15}\frac{61000 - 32768}{n\cdot2^i}\right\rceil = \left\lceil \sum_i^{15}\frac{28232}{32\cdot2^i}\right\rceil = 883$ probe-triplets.

The first requirement of the port exposing technique is that the web pages support persistent HTTP connections. In such connections, all requests are over the same (victim) connection and ensure it does not close. Persistent HTTP connections are the default configuration of apache servers and are also employed by many large web-servers (e.g., Facebook, Yahoo!, Google), but not all (e.g., live.com). We require persistent connections since otherwise, the connections opened in step 1 of this attack will close by step 2. 

The second requirement of the port exposing technique is that the connection will not enable the selective acknowledgment (SACK) TCP option. This option allows the receiver specify exactly which packet arrives when sending an acknowledgment. The three duplicate acknowledgments sent in our attack are identical (specify the same SACK option), i.e., indicate that no server packet had arrived between these packets. In this case the server, who receives these acknowledgments, may treat these as a single duplicate acknowledgment and will not reduce the rate. We also specify this mechanism as possible defense in Section \ref{Injection:Defenses} (Defenses).

\begin{figure}
  \begin{center}
    \includegraphics[width=0.5\textwidth]{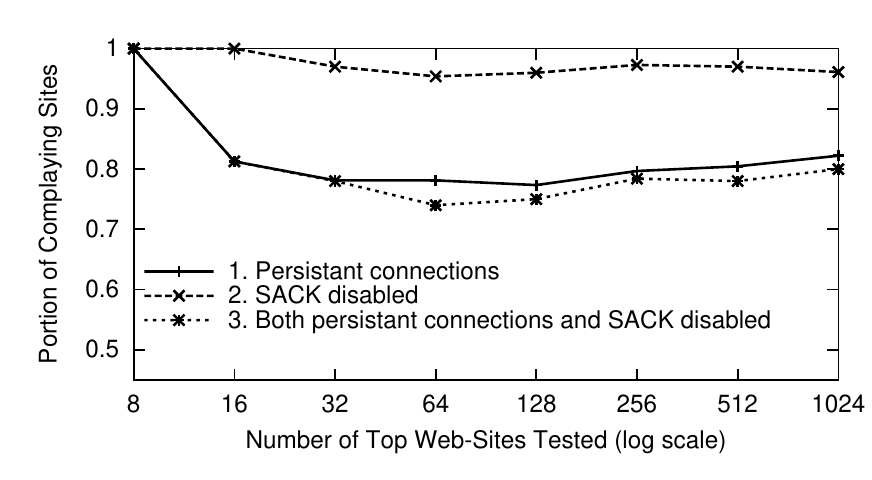}
  \end{center}
  \caption{Portion of popular websites that comply with attack requirements.}
    \label{fig:sitesupport}
\end{figure}

Figure \ref{fig:sitesupport} shows the portion of 1000 web-servers that comply with either and both these requirements.

\ignore{
The third requirement, is that the spoofed probe packets arrive at the client. We assume that \mal\ is capable of spoofing, and hence, that there is no spoofed packet filtering at the attacker's end. However, this attack requires that such filtering is not performed at the client's end. In particular, we require that if a firewall that filters packets by sequence number is deployed
Firewall allowing sequence numbers...
}

\subsection{Empirical Evaluation}
We evaluated the port exposure attack described in this section. Our initial results show that the technique has high success rate (over 70\%). We are currecntly working on further improving theses results. In the following versions of this paper we will include details and graphs on our experiments.

\section{Search for Server's Sequence Number} \label{serverseq}

In this section we present the first phase of ``inject and observe'', a technique for learning the sequence numbers used in the victim-connection; we follow the high level design presented in Section \ref{injectandobservehigh}. At the end of this phase \mal\ learns the server's sequence number; this allows her to send data to the client impersonating as the server. We assume that \mal\ has the parameters of the victim-connection, in particular, that she identified the client port; e.g., by executing the technique described in the previous section (or other methods, see Table \ref{Tbl:blocks}). 

 Subsection \ref{http} provides required background, explaining how browsers handle HTTP responses that they receive. Subsection \ref{serversnum} describes our search technique. In the following section we present the second phase of ``inject and observe'', where \mal\ learns the client's sequence number; in that section we provide an empirical evaluation for the entire procedure.

\subsection{Browser HTTP Request/Response Handling} \label{http}

Browsers support persistent HTTP connections and request pipelining since HTTP 1.1 (\cite{rfc2616}). These allow sending of multiple requests to the same server in pipeline over a single TCP connection. The responses from the server are buffered and ordered according to the browser's flow control window, denoted by \wnd\ (allocated per-connection). This window is important to our attack; it specifies the range of disordered bytes that the TCP receiver (browser) can keep, its size is typically $2^{16}$. Data bytes whose corresponding sequence number is in \cwnd\ are buffered in a receive-queue for the application (browser), other arriving data is discarded.

The browser keeps a queue for transmitted HTTP requests and waits for a response if that queue is not empty; i.e., reads from the TCP `received-buffer' until reading a full response. The response's HTTP layer is parsed and its encapsulating data is then embedded in the web page, the request is removed from the queue. This process continues until there are no more requests awaiting reply from the particular connection.

Unfortunately, the HTTP specification does not tell the appropriate action when a response does not have a valid HTTP header (i.e., parsing fails). The de-facto standard, used by current versions of Internet Explorer, Firefox and Chrome (and possibly other browsers as well) encapsulates all data that is available for reading, i.e., sequence of received bytes starting from the low bound of \cwnd, under the following single HTTP header:

\begin{verbatim}
 HTTP/1.1 200 OK
 Content-Type: text/html; charset=us-ascii
 Content-Length: available-data-size
\end{verbatim}

Browsers do not break the existing TCP connection in this case and continue processing requests over it. We believe that this behavior, which essentially displays content data as plain text, was introduced in early days of the Internet to handle misbehaving servers and was mimicked in modern implementations. The following subsection explains how this behavior is exploited to learn the server's sequence number.

\subsection{Inject and Observe} \label{serversnum}
\begin{figure}
  \begin{center}
    \includegraphics[width=0.5\textwidth]{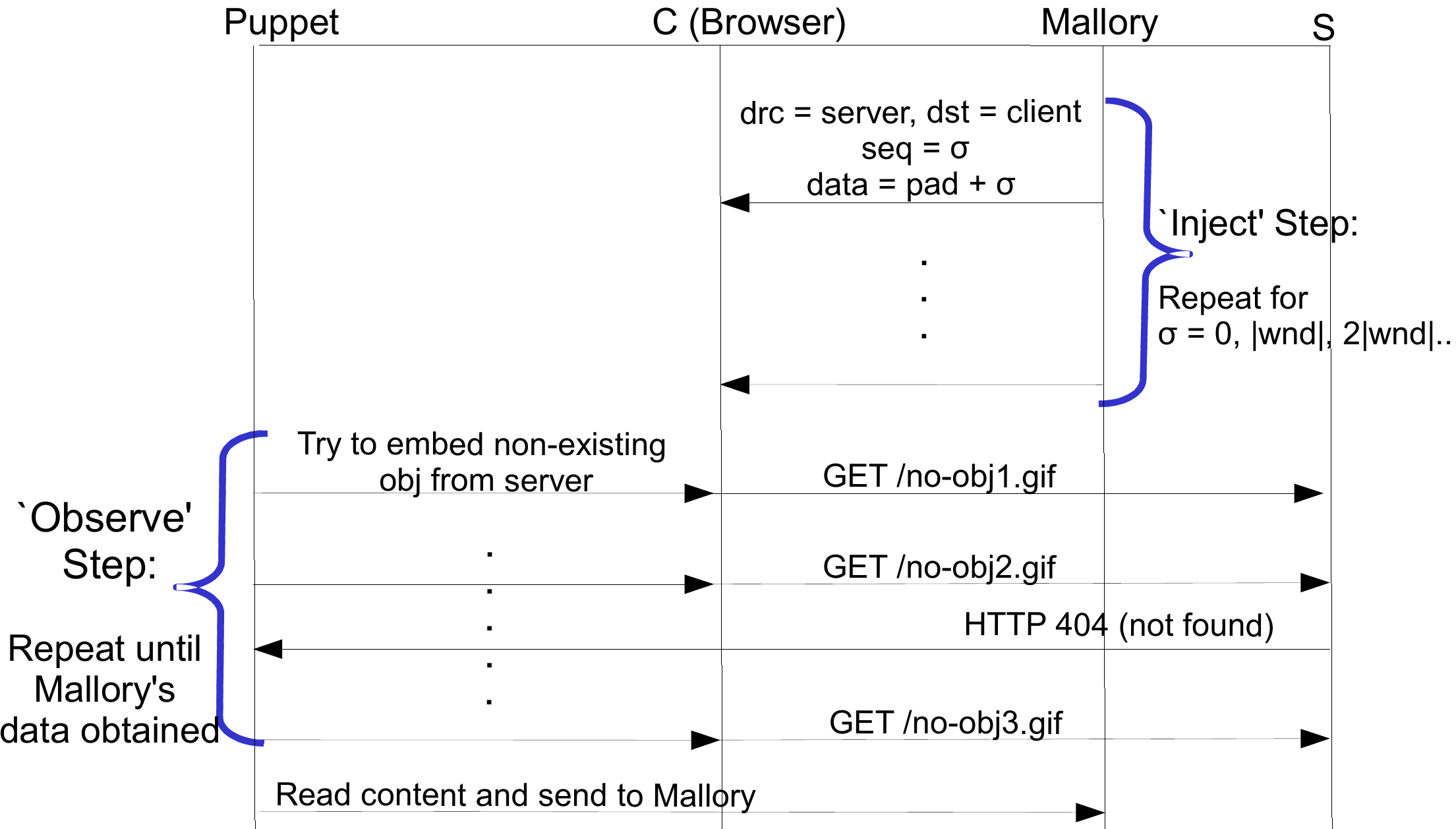}
  \end{center}
  \caption{Server Sequence Number Learning Technique.}
    \label{fig:injectandobserve}
\end{figure}

In this subsection we present the server sequence number learning technique which is illustrated in Figure \ref{fig:injectandobserve}. The search has two steps `Inject' and `Observe'.

In the Inject step, illustrated by the top packets in Figure \ref{fig:injectandobserve}, \mal\ sends spoofed packets that appear to belong to the victim connection. Each of these packets specifies a sequence number that is \wndsize\ greater than the previous one. \mal\ writes (as part of the TCP data) the packet's sequence number and prependers it with padding. 

All the packets that \mal\ sends except for one are immediately discarded by \wini\ since the sequence number that they specify is out of \wnd\ boundaries. For simplicity, we assume that the packet that is saved in \wini's buffer does not begin at the lowest offset in \wnd\footnote{The probability that this event occurs is rather small: $\frac{\textit{packet-length}}{\text{\wndsize}} \leq \frac{1500}{2^{16}} \approx 0.02$.}. \wini\ responds to every packet that \mal\ sends with a duplicate Ack; this will cause connection slow down (but not breakdown) as we exploited in the previous section, however, this does not effect the sequence number learning process. 

There is an additional validation that is performed on the acknowledgment number specified in received packets. It verifies that the packet does not specify a future Ack; we refer (and exploit) to this condition extensively in the following section. In this section the attacker is therefore required to send two packets for each sequence number, one specifies $\textit{Ack} = \alpha$ and the other specifies $\textit{Ack} = \alpha + 2^{31}$; ensuring that the Ack in one packet is valid.

\begin{figure}
  \begin{center}
    \includegraphics[width=0.4\textwidth]{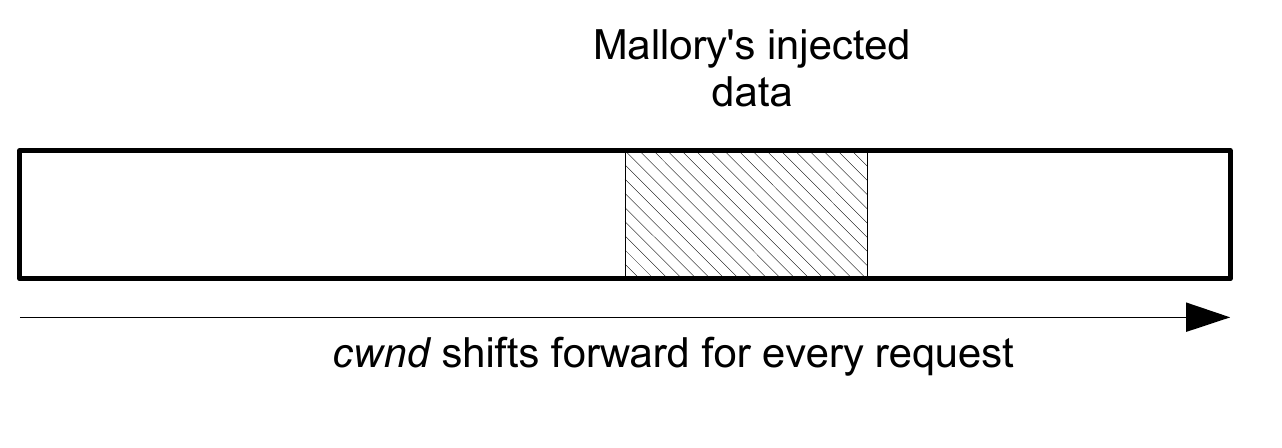}
  \end{center}
  \caption{The state of \wnd\ after `Inject' step.}
    \label{fig:flowcontrolwin1}
\end{figure}

After this step, \wini's victim-connection \wnd\ is as illustrated in Figure \ref{fig:flowcontrolwin1}. 

In the following Observe step, the puppet requests objects from \lin. It ensures that there is always at least one request waiting for reply in the browser's queue; this by generating two initial requests and sending a new request when a response arrives. See illustration in Figure \ref{fig:injectandobserve}. The reason that one request must be in queued is that when there are no pending requests, some browsers clear the receive buffer (those will discard the injected data); this behavior was verified on Mozilla Firefox and Google Chrome. 

Each response that arrives at \wini\ shifts \cwnd\ forward; eventually, the injected data that was saved in the previous step will be used as a response. Some of the last server's response might overlap the injected data. The TCP specification determines that in this case the more recent data is used, this essentially means that some of the injected data will be lost. The padding that prepends the sequence number in the response is dispensable data, that is of the server's response size; it ensures that the sequence number value is not overrun by a server's response. 

The requests that the puppet sends are for arbitrary objects that will yield an HTTP 404 response (object not found). The reason that we prefer receiving such responses is that they are relatively short which reduces the amount of data that the adversary sends. Furthermore, using such requests makes the attack generic: \mal\ need not identify specific objects to obtain for every remote server she uses.

When the puppet reads a response that does not specify `object not found' it sends the response to \mal\ who finds the sequence number; adding the length of the packet she sends she identifies the next sequence number. \footnote{Since the puppet (Java Script) runs in context of \mal\ domain, they can communicate without restrictions.}

\begin{figure}
  \begin{center}
    \includegraphics[width=0.4\textwidth]{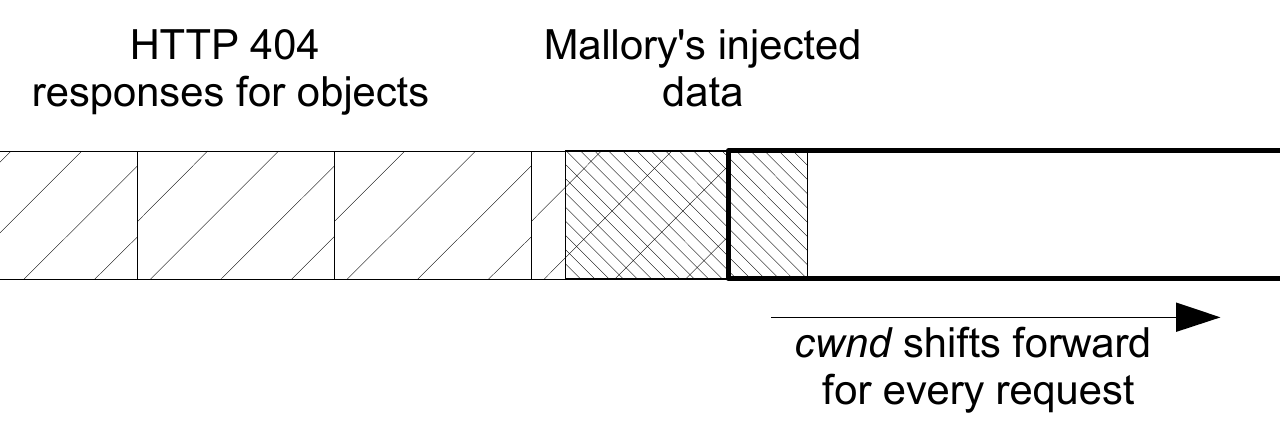}
  \end{center}
  \caption{The state of \wnd\ during the `Observe' step.}
    \label{fig:flowcontrolwin2}
\end{figure}

The technique that we presented in this subsection requires the adversary to send a number of packets that is linear to the number of sequence numbers. Specifically, the adversary sends during the `Inject' step $2\frac{2^{32}}{2^{16}} = 2^{17}$ packets.


\section{Search for Client's Sequence Number}\label{Injection:AckEsposure}
In this section we present the second phase of the `inject and observe' technique. At the end of this phase \mal\ learns the client's sequence number; this allows her to send data to the server impersonating as the client. We show how \mal\ can perform a binary search for the Ack number that the client expects which equals the sequence number that the client will next send. This step assumes that \mal\ has the parameters of the victim-connection and learned the server's sequence number. Our technique also extends the one presented in \cite{snptcp} which detects the server's sequence number.

Subsection \ref{tcprecv} provides necessary background on TCP's packet processing, Subsection \ref{clientsnum} presents the binary search technique.

\subsection{TCP Receiver Packet Handling} \label{tcprecv}

\begin{figure}
	\centering
		\includegraphics[width=0.481\textwidth]{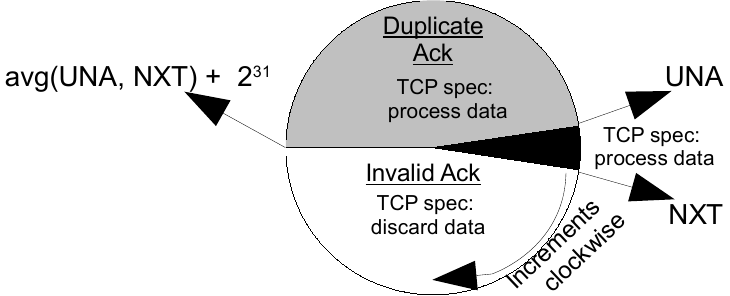}
	\caption{Ack Number Map. UNA is the lowest unacknowledged sequence number, NXT is the next sequence number that \wini\ will send. \ignore{Acknowledgments for data in the {\em gray} area are considered duplicates, acknowledgments for data in the {\em white} area are for unsent data, i.e., invalid. In the {\em black} area are sequence numbers for sent un-Acked data. }The 32-bit Ack field is cyclic.}
	\label{fig:AckMap}
\end{figure}

In the technique that we describe in the following subsection \mal\ learns whether some acknowledgment number is above or below the sequence number that the client will next use, denoted by NXT. Since the sequence number is a cyclic field, when we write above or below, we refer to the drawing in Figure \ref{fig:AckMap}. The figure illustrates a division of the Ack field values into three intervals: in the gray area (sent and Acked) and the black area (sent, but not Acked) are numbers below NXT. In the white area are sequence numbers above NXT. 

The test is derived from an observation from the TCP specification \cite{rfc793} (Section 3.9, page 72). The relevant statement refers to an acknowledgment packet that carries data and contains a valid sequence number; i.e., success in learning the server sequence number is required to initiate this phase. The specification distinguishes between two cases regarding the acknowledgment number in the packet. 

{\bf Case 1:} the packet contains a duplicate Ack (gray area in Figure \ref{fig:AckMap}), or acknowledges data that was sent, but not already acknowledged (black area in Figure \ref{fig:AckMap}). In this case, the recipient continues processing the packet regularly (see \cite{rfc793}) and the application will receive the data bytes.

{\bf Case 2:} In the complementary case, that the acknowledgment number is for data that was not yet sent (white area in Figure \ref{fig:AckMap}), the recipient discards the packet.

\subsection{Inject and Observe} \label{clientsnum}
We now describe the Inject and Observe technique to learn the Ack value. Each iteration of Inject and Observe eliminates half of the possible values to the Ack field. We initiate this phase assuming that all Ack values are possible. Let \textit{ack-low} denote the lowest possible value for the acknowledgment field and let $l$ denote the number of possible values for the field; initially, $\textit{ack-low} = 0, l = 2^{32}$. 

\begin{figure}
	\centering
		\includegraphics[width=0.48\textwidth]{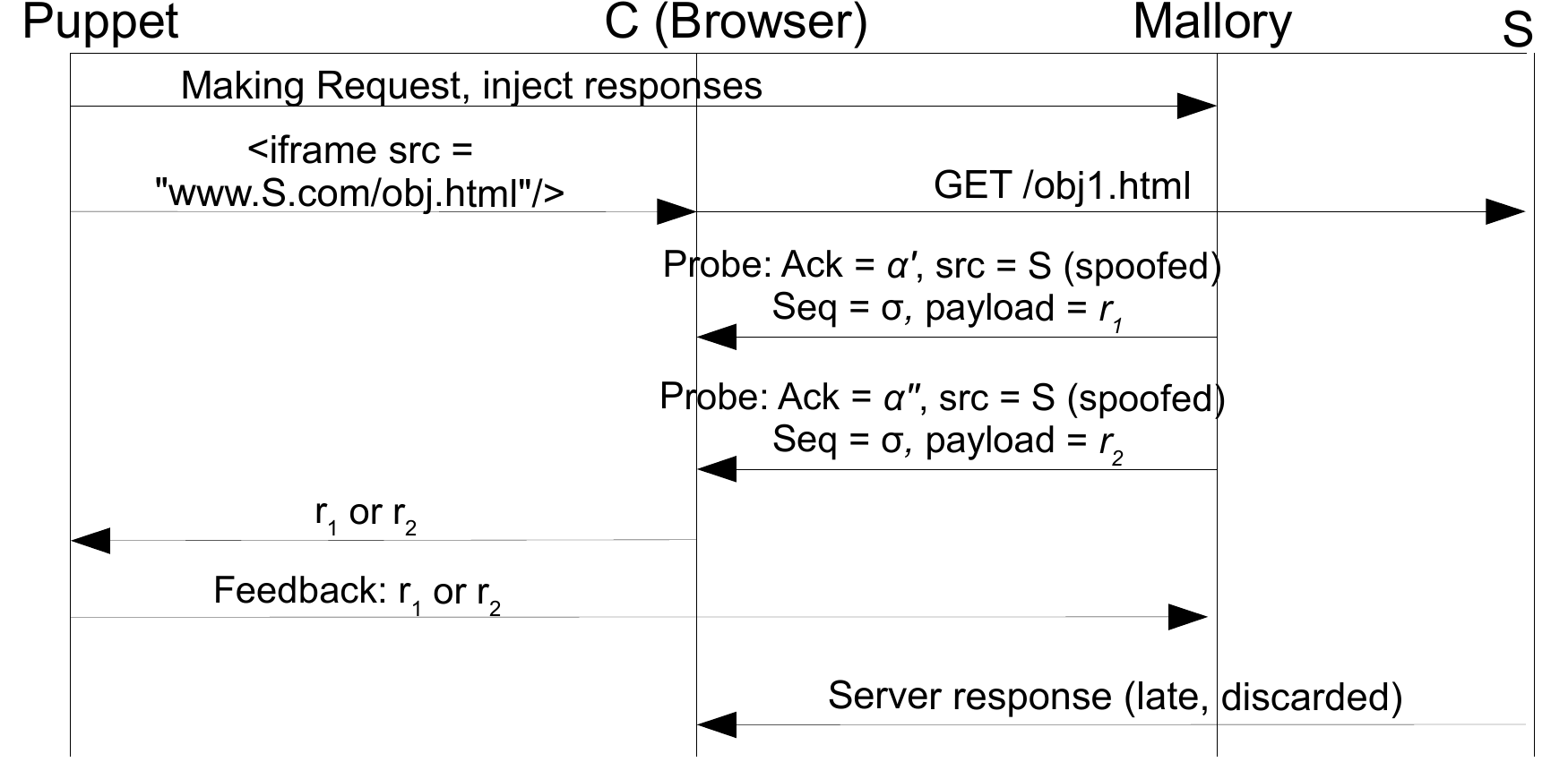}
	\caption{Client Sequence Number Learning Technique.}
	\label{fig:AckTest}
\end{figure}

In every iteration of the search the puppet requests a HTML page and the adversary provides a response injected to the TCP stream by specifying the server's sequence number learned in the previous step, denoted by $\sigma$. The adversary sends two HTTP responses that specify different pages: $r_1$ and $r_2$. The response packets specify the acknowledgment numbers $\alpha' = \textit{ack-low}, \alpha'' = \textit{ack-low} + \frac{1}{2}l$. According to the observation above, one of these packets specifies an Ack that is not above the true Ack value that the recipient expects (NXT) and will be discarded. Therefore, the puppet receives either $r_1$ or $r_2$ and will report the result to \mal. Figure \ref{fig:AckTest} illustrates a search iteration.

Let $\alpha$ be the ack number of the packet that arrive at the client (identified by puppet's feedback). After each iteration \mal\ updates the parameters $\textit{ack-low}, l$ and $\sigma$: $\textit{ack-low} \leftarrow \alpha, l \leftarrow \frac{l}{2}$ and $\sigma \leftarrow \sigma + \textit{response-size}$ in order to account for the injected data.

\subsubsection{Binary Search for $\alpha$} \label{Injection:SimpleBinary}

The Inject and Observe technique above discovers the sequence number in a binary search methodology. The `before' and `after' terminology used in the TCP specification (illustrated in Figure \ref{fig:AckMap}) divides the possible values of the Ack field to similar-sized areas: the gray and white areas in Figure \ref{fig:AckMap} are of equal size, and the black area (sent bytes without acknowledgment) is usually relatively small. Therefore, each iteration of the attack eliminates approximately half the possible values for the field. This allows \mal\ to perform a binary search for NXT; each time eliminating approximately half the possible numbers. The 32-bit length of the Ack field implies that there are $32$ iterations.

\subsection{Empirical Evaluation}
We evaluated the inject and observe technique described in this section. Our initial results, which will be included in following versions of this paper, show that the technique has considerable success rate as a standalone componenet and when combined with the port exposure technique described in this paper (over 30\% and 20\% respectively). We are currecntly working on improving theses results.

\section{Denial-of-Service Exploits} \label{exploits:DoS}
We next describe possible exploits of long-lived-connection injection attacks to disrupt network communication. Namely, we show how attackers can build on off-path injection techniques, to launch formidable Distributed Denial-of-Service (DDoS) attacks. As before, we consider an off-path IP-spoofing attacker, who also controls a significant number of low-privileged malwares or `puppets', i.e., scripts running in browsers of unsuspecting users. 

Performing typical denial of service attacks with weak malicious agents is challenging, as we motivated in Subsection \ref{dosandother}.

\subsection{Off-Path Optimistic Ack Attack} 
We first describe {\em Off-path Optimistic Ack}; this is a variant of the Optimistic Ack DoS attack  \cite{SBB05:OptAck}. These attacks cheat TCP's congestion control mechanism, causing senders to believe that most of the data they sent was already received, and hence that they can send more data (congestion window is not full), and also to increase the size of the congestion window and as a result, the sending rate. This can result in huge amplification factors, see \cite{SBB05:OptAck}, unless servers use per-connection bandwidth quotas or use other defenses. 
 
In both Optimistic Ack attacks (ours and the original \cite{SBB05:OptAck}), the attacker sends to the server acknowledgment packets (Acks), as if the client received all packets sent by the server (although packets are still in transit or even lost). As a result, the server continues sending information, with increasing rates. 

In the original attack \cite{SBB05:OptAck}, the client must run malware, with ability to send `raw IP' packets (i.e., not according to the TCP specifications). This is a significant requirement; recent operating systems make it harder for malware to obtain such ability, which may only be available to privileged (i.e., `root') programs. Furthermore, the original attack may be blocked by a firewall on the client side, by detecting the unusual high rate. Note that since by requiring only puppets, attacker is more likely to control enough clients to succeed in the attack, in spite of server-side countermeasures such as per-connection quotas. 

\subsubsection{Off-Path Variant Attack Process}
A TCP injection attack allows an attacker to perform an off-path variant of the Optimistic Ack attack as follows. The attacker only needs a puppet on the client machine to open the TCP connection with the victim server and learn the client port and sequence numbers. Following this, the puppet requests some large object from the server and is no longer needed; the attacker sends Ack packets as done by the client in the original attack, and - if not using SSL/TLS - the attacker can even send new request(s) if needed (instead of the client). Note that even if the client's firewall detects the attack and begins blocking packets on this connection from both directions, this does not help since we provide the Acks to the server from the off-path attacker. Furthermore, RFC-compliant RST packets that the firewall (or client) may send, would be out of the server's flow control window and hence ignored, and would not tear the connection. Server-induced verifications, by intentionally dropping or reordering packets periodically, as suggested in \cite{SBB05:OptAck}, seem one of the best (or only) defenses.

\subsection{Off-Path Ack-Storm DoS Attack}

In the original Ack-Storm DoS attack \cite{AH11:Ack}, the attacker needs to have some ability to eavesdrop on packets. From these packets, the attacker learns the TCP parameters (IP addresses, ports, and sequence numbers). Using these, the attacker sends two spoofed data packets, one to each of the two ends of the connection. According to the TCP specification, and in most TCP implementations, upon receiving an Ack for data that was not yet sent, TCP sends back a `duplicate Ack' - i.e., resends the previously sent Ack. As a result of receiving the pair of spoofed data packets, one at each peer, both peers begin sending acknowledgment packets to each other. Since these Acks acknowledge data which was actually sent by the attacker, not by the peer, then each of these Acks will only result in another duplicate Ack returned, and this `ping-pong' process will continue indefinitely. 

The attacker can send additional data packets to the peers, causing additional ping-pong exchanges, quickly filling-up the channel capacity. This causes increased load on the networks; the fact that the packets involved in the attack are very short (just Acks),  makes the load on routers and switches even higher. Eventually, this causes packet losses, and legitimate TCP connections sharing the same links significantly reduce their rate. 

\subsubsection{Off-Path Variant Attack Process}
The off-path Ack-Storm DoS Attack works exactly like the regular Ack-Storm DoS Attack, except for employing the injection techniques to allow the off-path attacker to learn the TCP parameters. Hence, the attacker can run this attack without requiring the ability to eavesdrop. 

\subsection{Empirical Evaluation}
\begin{figure}
  \begin{center}
    \includegraphics[width=0.35\textwidth]{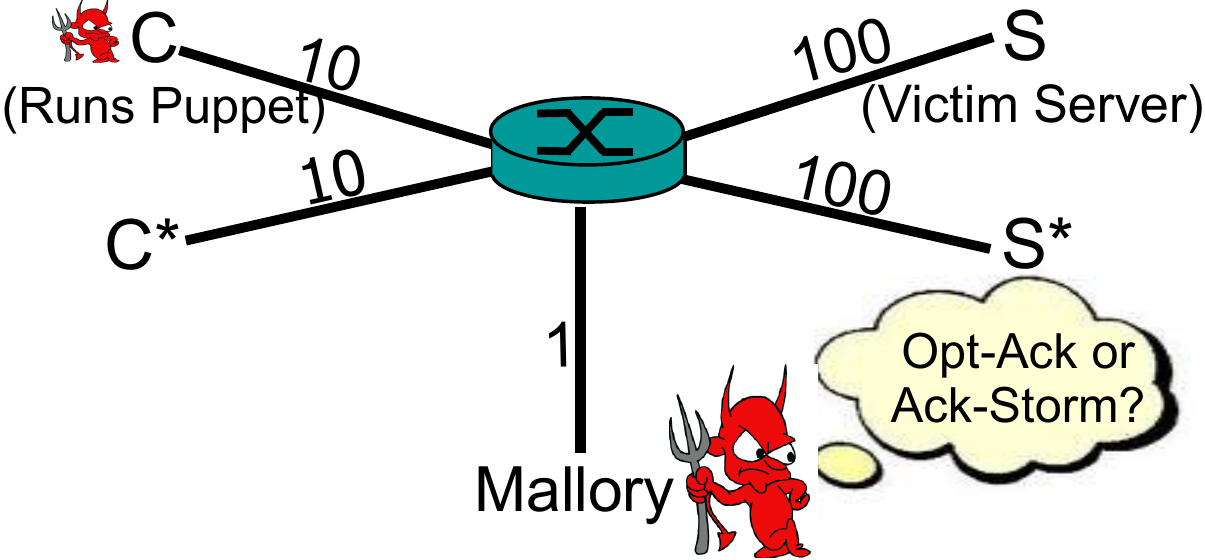}
  \end{center}
  \caption{Network topology for testing the DoS attacks. Each link specifies its capacity in mbps.}
  \label{fig:dosmodel}
\end{figure}

We used the topology illustrated in Figure \ref{fig:dosmodel} to test the off-path denial of service attacks that we presented. In our tests we assume that \mal\ runs a puppet on \wini\ and can inject data to the TCP connection between \wini\ and \lin\ (a connection that \mal\ caused \wini\ to establish).

We measure the degradation of service that these attacks cause to other legitimate connections. We consider different round-trip times (RTTs) for the legitimate connections that we measure: the longer the RTT is, the greater the congestion windows. Since every loss halves the congestion window, a more significant effect is observed when RTT is high. Furthermore, a higher RTT implies that it will take more time for the legitimate TCP sender to detect a packet loss and retransmit. We compare the effect of these attacks to line 1 in Figure \ref{fig:AckStromOptAck} (base-line) which illustrates the time it takes \wini* to receive a 50MB file from \lin* under normal conditions.

\subsubsection{Off-Path Optimistic Ack Evaluation}
\begin{figure}
    \includegraphics[width=0.45\textwidth]{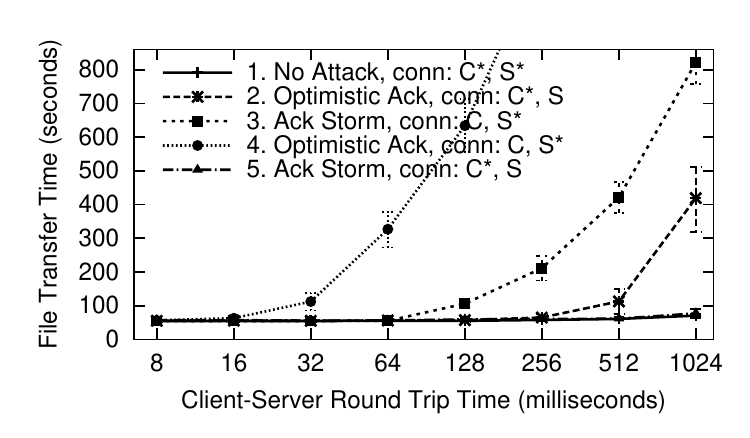}
  \caption{Evaluation of Ack-Storm and Opt-Ack DoS attacks. The legend indicates for each line the type of attack and the peers in the legitimate connection (that \mal\ degrades). Measurements are the average of 50 runs, error-bars mark the standard deviations.}
  \label{fig:AckStromOptAck}
\end{figure}

In this attack we aimed to clog the \lin's link: \mal\ uses \wini\ to request some large file from \lin\ and then performs the Optimistic Ack attack, persuading \lin\ to send data to \wini\ at high a rate. We evaluated the effect of this attack by measuring the degradation of service in a connection that \lin\ has with some other client \wini* who tries to download a 50MB file\fullppr{ (see network topology in Figure \ref{fig:dosmodel})}. Lines 1 and 2 in Figure \ref{fig:AckStromOptAck} illustrate our results. Notice the significant difference in attacker and server link capacities; the amplification ratio measured in this attack is $78$ (for every byte that \mal\ sends, \lin\ sends approximately 78 bytes). Furthermore, the attack also clogs \wini's link, as shown by line 4.

\subsubsection{Off-Path Ack-Storm Evaluation}
We use the Ack-Storm attack to congest \wini's link and measure the effect on a different connection that he has with some other server \lin*. In order to congest the link, \mal\ creates a new Ack `ping-pong' every 100 ms; to create each ping-pong \mal\ sends only two short (40B) packets. In the connection with \lin*, \wini\ tries to download a 50MB file (similar to the previous experiment). Lines 1 and 3 compare the file transfer time at a normal time, to that when the Ack-Storm attack takes place. This attack requires much less effort and lower bandwidth than the Opt-Ack attack, i.e., has much higher amplification ratio. However, the Ack-Storm attack is limited by the client bandwidth (which is typically lower than the server's); this limitation is illustrated by line 5 which is very similar to line 1 (normal conditions). 

\section{Off-path Coremelt Attack} \label{coremelt}
The two DoS attacks which we described in Section \ref{exploits:DoS} can be launched using only puppets and have high amplification ratio, it follows that even relatively weak attackers may be able to cause large amounts of traffic from many clients spread around the network. This can cause high load on servers, routers and links. 

In particular, by choosing well the pairs of clients and servers between which the attacks are launched, the attackers can cause huge amounts of traffic to flow over specific `victim' backbone routers and links. These backbone networks, connecting large core ISPs (autonomous systems), have very high capacities; by sending enough traffic to a particular destination, attacker can cause queuing and losses in the connecting router. As a result, Internet connectivity may break - first for TCP connections and then even for UDP applications. 

We use the term {\em Off-path Coremelt Attack} for the resulting attack on core Internet connectivity, since it is an off-path variant of the Coremelt attack \cite{conf/esorics/StuderP09}. The Coremelt attack uses a large {\em botnet}, sending large amounts of traffic between pairs of the bots, with the pairs chosen intentionally so that huge amounts of traffic will flow over specific victim link/router. As shown by simulations in \cite{conf/esorics/StuderP09}, this can result in congesting the victim link/router and breaking connectivity in the Network. We discuss the attack and analyse it, presenting an interesting optimization problem for the attacker; notice that similar analysis is applicable to the original Coremelt (and was not presented so far).

\ignore{
Our evaluation of the off-path Coremelt attack will includes use simulations, extending on these of \cite{conf/esorics/StuderP09} (and comparing to their results); while we already have interesting results, we feel that this part of our study still is not ready for publication, and furthermore, there is not enough space in this submission to justify including these.}

\subsection{Off-Path Variant Attack Process}
In the Off-path Coremelt attack, the attacker congests a core link or router; however, instead of depending on pairs of zombies (bots), here the attack just requires a puppet at one end. The other end of the connection is a legitimate server, and to cause huge amounts of traffic, the attacker uses one of the two off-path DoS attacks described above. 

This attack has three advantages compared to the original Coremelt attack: (1) controlling a large and correctly-dispersed set of puppets is easier than controlling a comparable set of zombies; (2) we only need to control one end of each connection (puppet), not both ends; and (3) since we use adversary-chosen servers, these can have very high bandwidth, higher than available to most bots.  

\subsection{Problem Statement and Adversarial Algorithm}
The off-path Coremelt attack involves a non-trivial optimization problem for the adversary, i.e., how to cause maximal damage using the resources available to her, namely, mainly, the bandwidth of the {\em attacker}, i.e., of the attacking node itself, as well as the capacities of its {\em puppets}. The damage is essentially measured by the number of disconnected $<v,d>$ pairs, where $v$ is a {\em victim client} and $d$ is a {\em destination} (server that the victim needs to access). We note that both Opt-Ack and Ack-Storm attacks, utilize the unwitting cooperation of server machines at the other end of the connections; while in principle, such unwitting servers are also a limited resource, we believe that in practice there are so many such servers that we can safely ignore this limitation, as other limitations would almost always be the bottleneck. 
 
This `real' optimization problem is quite complex; we present and analyze a version with several significant simplifications, which, nevertheless, results in a very effective attack, significantly improving compared to the `original' off-core attack \cite{conf/esorics/StuderP09}. More detailed modeling and analysis, and more efficient algorithms, may allow an even more effective attack, and remain a challenge for future research.

Intuitively, the goal of the attacker is to break connectivity between autonomous systems (ASs), by congesting links with excessive traffic. Notice that this ignores the possibility that it may be sometimes more effective to attack some intra-AS link, e.g., the `last mile' connection of a particular host (destination server, or even client); we focus on `core-melt-like' attacks, which focus on inter-AS links. Attackers can only further optimize by congesting intra-AS links, when it is more effective. 

We focus on significantly degrading the performance of TCP connections. TCP is the most widely-used transport protocol, and due to its congestion-control mechanisms, it is very sensitive to packet loss (as in a clogging attack); as explained in \cite{AH11:Ack}, a very low loss rate suffices to cause major degradation of TCP performance. In particular, sending roughly as many packets as the link capacity, surely ensures sufficient, even significant, loss rate. 

For clogging TCP connections, the attacker can utilize both off-path attacks (Ack-Storm and Opt-Ack), choosing effective mix of the two, based on the given topology of attacker puppets, victims and servers. The Ack-Storm attack requires the attacker to send new packets whenever on of the attack packets is lost; hence, it is only effective for low loss rates. Indeed, intuitively, the Opt-Ack attack seems much more effective, due to its huge amplification factor; in our measurements, we found average amplification factor of $\ampfact=78$, which is very impressive - although significantly less than the amplification reported by the original Opt-Ack research (and incomparable to the theoretical analysis presented there), see \cite{SBB05:OptAck}\footnote{We believe the difference between our results  and these of \cite{SBB05:OptAck}, may be mostly due to differences in the (default?) quotas and other restrictions, of the web servers used.}. For simplicity, in our analysis we simply assume that Opt-Ack attacks result in fixed bandwidth amplification of $\ampfact$, i.e., if the attacker dedicates bandwidth of $x$ bytes per second to Opt-Ack attack by sending it to a particular server $s$, injected into connection with some (attacking-puppet) client $c$, then the result is transmission of $\ampfact\cdot x$ bytes from $s$ to $c$. We ignore the (nominal) amount of traffic required for this attack by the client attacking-puppet $c$. 

In contrast, the Ack-Storm attack requires the puppets of the attacker to send one packet per each packet sent by the abused server at the other end of the connection. However, notice that Opt-Act amplification factor applies to packet sent by the attacker, not by the puppets; and with Ack-Storm, significant degradation is possible already with low loss rates, e.g., even 0.1\% by our measurements, which essentially implies amplification factor of $1000$ (for the packets sent by the attacker). Hence, for Opt-Ack attacks, the limiting factor is usually the capacity of the attacker (albeit amplified by $\ampfact$), and for Ack-Storm attacks, the limiting factor is usually the capacity of the attacking puppet (without amplification). 

Hence, the following simplification seems reasonable; we assume that the attacker bandwidth is essentially used to its maximal capacity for Opt-Ack attacks, and the  capacity of attacker-controlled puppets (puppets and \wzs) is used for Ack-Storm attacks, ignoring the small additional amount of attacker traffic required also for Ack-Storm attacks and the small amount of attacking-puppet capacity used for Opt-Ack attack. This allows us to design the attacker strategy in two separate steps: first, we analyze the best allocation of the attacker bandwidth to Opt-Ack attacks; and then we analyze the best allocation of the capacity of the attacker puppets, to Ack-Storm attacks. 

For both allocations, we assume that the adversary knows the topology and routing of the AS-level Internet routing network, for example, as provided by route views \cite{routeviews}.
We model the topology of this network as a graph $G=(A,E)$, where $A$ is the set of Autonomous Systems (ASs), and $E$ is the set of links (edges) between ASs. For every link $(f,t)\in E$, let $c_{(f,t)}\in [0,\infty)$ be the {\em capacity} of $(f,t)$. As explained above, the goal of the adversary is to {\em clog} one or more links, causing maximal damage, i.e., disconnecting the maximal number of $<v,d>$ pairs; clogging link $(f,t)\in E$ essentially means causing $c_{(f,t)}$ or more traffic to be sent over link $(f,t)$. 

We assume that both attacker traffic and legitimate traffic, use fixed, known routes; for any given source AS $s\in A$ and destination AS $d\in A$, let the {\em route} from $s$ to $d$, denoted $r(s,d)=\{(f_i,t_i)\}_{i=1}^l$, be a sequence of some number $l$ of edges (in $E$), such that $f_1=s$, $t_l=d$, and for every $i:1\leq i < l$ holds $t_i=f_{i+1}$. 

Each of the relevant entities - victims, destinations, puppets and the attacker herself - reside in one of the ASs in $A$. For simplicity, we assume a single attacker, in a specific AS, with bandwidth $\atkbw_A$; without loss of generality, we assume this is AS number $666$. In contrast, victims, destinations and attacker puppets are spread among the ASs. For simplicity, we just consider two (potentially overlapping) subsets of $A$, denoted $A_v$ and $A_d$, and  representing the ASs containing victims and destinations, respectively (i.e., we ignore the issue of {\em how many} victims and destinations are in each AS). Finally, for every AS $a\in A$, we let $\atkbw_a$ denote the (total) bandwidth of attacking puppets in AS $a$. 

For simplicity, we assume that every AS contains sufficient number of servers that the attacker can use in Opt-Ack and Ack-Storm attacks, to generate traffic. Let $\atkbw_{a,s}$ denote the amount of traffic used in Ack-Storm attack, from attacking puppets in AS $a$ to the servers in AS $s$, and let $\atkbw_{A,c,s}$ be the amount of Opt-Ack attack traffic that the attacker sends to server in AS $s$, for a connection with client in AS $c$. Clearly, we have the following restrictions, for every $a\in A$: 
\begin{eqnarray}
\atkbw_a & \geq &\sum_s \atkbw_{a,s} \\
\atkbw_A & \geq & \sum_s \sum_c \atkbw_{A,c,s} 
\end{eqnarray}

We next require that the total traffic on each link, including Opt-Ack traffic (from attacker to server, and then, amplified by $\ampfact$, from server to client) and Ack-Storm traffic (from client to server and back), is not greater than the link capacity, namely, for every $(f,t)\in E$, we have: 

\begin{eqnarray}
c(f,t) & \geq & \sum_{<s,c>|(f,t)\in r(s,c)} \ampfact \cdot \atkbw_{A,c,s} + \nonumber \\
       &      & + \sum_{s| (f,t)\in r(A,s)} \sum_c \atkbw_{A,c,s} + \nonumber \\
       &      & + \sum_{a,s\in A|(f,t)\in r(a,s)} \big( \atkbw_{a,s}+\atkbw_{s,a} \big)
\label{eq:trans}
\end{eqnarray}

We denote the right term in Equation \ref{eq:trans} by $T(f,t)$, i.e., the transmitted communication over the edge $(f,t)$.
Finally, we want to maximize the number disconnected pairs $<v,d>$. We use the same criteria as the original Coremelt attack and consider a pair $<v,d>$ disconnected if there exists a link on the route between them that is loaded up to its capacity. Denote by $x_{<v,d>}$ an indicator variable for clogged routes:

\begin{eqnarray} \label{xs}
x_{<v,d>} = 1, \text{ If: } \exists\text{ }(f,t)\in r(v,d)\text{ }|\text{ }c(f,t) = T(f,t) \nonumber\\
x_{<v,d>} = 0, \text{ Otherwise\hspace{118pt}}
\end{eqnarray}

Our goal function is:
\begin{eqnarray} \label{goal}
\max \sum_{<v,d>} x_{<v,d>}
\end{eqnarray}

We have stated the off-path Coremelt attack as a maximization problem and would like to use linear programming to solve it, i.e., find the amount of data sent by each attacking node to every server (denoted by $\alpha$ variables). However, the indicator variables defined in Equation \ref{xs} cannot be represented as linear terms since the quantitative `$\exists$' is not linear. We solve a relax version of the maximization problem where the target function and constraints are as the regular definition of the problem, but the definition of the indicator variables change. 

Let $\textit{Cut}(G)$ denote the minimal cut of the network graph $G$ with respect to the sets of victims and destinations. For every pair $<v,d>$, denote by $\textit{target}(v,d)$ the edge with minimal capacity such that $\textit{target}(v,d) \in r(v,d)$ and $\textit{target}(v,d) \in \textit{Cut}(G)$. Define:

\begin{eqnarray}
x_{<v,d>} = 1, \text{ If: } c(\textit{target}(v,d)) = T(\textit{target}(v,d)) \nonumber\\
x_{<v,d>} = 0, \text{ Otherwise\hspace{72pt}}
\end{eqnarray}

In the relaxed version we target specific links in the route between victims and destinations, those that belong to the minimal cut. This restriction allows stating the attack as a linear programming maximization problem: the goal function is Equation \ref{goal} and the constraints are given by Equations 1 - 3.


\section{Defense Mechanisms} \label{Injection:Defenses}
The attacks presented in this paper relay on successful exposure the client port and sequence numbers. In this section we propose remedies, mitigating the attack vectors considered in this paper. We describe both immediate patches that can be deployed with only minor modifications longer term remedies that require modifications to existing implementations and deployed devices. The remedies purposed below are of two types, those deployed at the client-end, and those deployed at the server-end. Each mechanism blocks the attack even if the other peer is vulnerable. 

\subsection{Server-End Defense}

\subsubsection{Mitigation for Client Port Exposure}
A server can mitigate the port exposure technique described in Section \ref{clientport} by enabling the selective Ack (SACK) TCP option. This option is supported by most modern client operating systems (including Linux) and is usually advertised in the client's SYN packet. However, we found that web-server rarely enable this option which is therefore typically not employed (see measurements in Figure \ref{fig:sitesupport}). When the SACK option is employed, every Ack that the client sends to the server specifies the sequence numbers corresponding to data that was received. This allows the sender, in case that one of his packets is lost, to identify and resend only the lost packet. 

In our attack, the three duplicate Acks that the client sends result from three sequential (spoofed) packets that the adversary sends. The SACK field in these acknowledgments is identical and identifies that these packets did not originate from three server-sent packets (since otherwise, the SACK field in every duplicate Ack would have increased for every packet). The server should handle such multiple duplicate Acks with identical SACK fields as a single duplicate Ack. The mechanism is already implemented in Linux (but typically disabled).

\subsubsection{Mitigation for Sequence Number Exposure}
The techniques presented in Sections \ref{serverseq}-\ref{Injection:AckEsposure} for learning the client-server sequence numbers inject data to the TCP stream; injected data is then observed by the puppet who feedbacks the attacker. In order to ensure data integrity, cryptographic mechanisms should be deployed; i.e., servers should use SSL/TLS instead of relying on randomized initial sequence numbers for authentication.

\subsection{Client-End Defense}
\subsubsection{Mitigation for Client Port Exposure}
Clients can modify the port selection algorithm used by the operating system. Of the three remaining algorithms suggested in \cite{rfc6056}, algorithm four is closely related to the Simple Hash-Based Port Selection algorithm exploited in this paper and might be vulnerable to similar attack. Hence other algorithms are preferred. 

However, in many cases modification of port allocation algorithm on the client machine does not suffice to prevent the attack. Many clients are connected to the Internet via NAT devices; these middle-boxes allocate each connection an external port and typically run an embedded version of Linux, i.e., use a vulnerable algorithm. Similar modifications are required to these devices as well.

\subsubsection{Mitigation for Sequence Number Exposure}
The server sequence number exposure attack that we presented in this work exploits de-facto standard browser behavior which is not required by standard: display corrupt responses to the user. Browsers can modify this behavior and in case that a response does not pass HTTP parsing, send a TCP reset to the server and close the connection. This modification conforms with the HTTP standard and protects the user from the attack vector considered in this paper. 


\ignore{as for the current version of Linux, there is no easily configurable alternate algorithm implemented in the system. Furthermore, in many cases,}

\ignore{

We next consider a different approach, allowing the browser to validate the responses it receives; this will protect against the data integrity attacks in Section \ref{exploiting:S} where the attacker `feeds' the browser with a fake response. The idea is that the browser adds a random identifying header to each HTTP request. The HTTP headers that the client specified are echoed in the server response and allow the client to validate the data he receives. This header will make injection much difficult since the attacker does not receive the identifier; she would need to inject her data in the middle of the server's response. This would also protect against other attacks such as Klein's response smuggling \cite{}.
}



\section{Conclusions and Future Work} \label{conclusions}
In this work present and evaluate a  new technique for off-path TCP injections. We show that a common implementation of client port randomization is vulnerable to off-path prediction attack. We investigate known DoS attacks in a new setting, one where the adversary is able to inject data into the TCP connection and show that this allows their deployment under weaker assumptions: puppets instead of bots, off-path instead of on-path attackers. This work continues the line of recent works on TCP injections, showing that the folklore belief that TCP is immune to off-path adversaries is incorrect. We motivates deployment of cryptographic protocols to protect communication, such as SSL/TLS and IPsec and believe that a more significant portion of servers should use these defenses, even if communication is not sensitive. 

This work leaves several directions for future work. The first is to test the off-path Coremelt attack with simulations, extending these of \cite{conf/esorics/StuderP09} (and comparing to their results). A second research direction is analyzing the security of the remaining three port randomization algorithms suggested in \cite{rfc6056}. In particular, we believe that the fourth algorithm suggested in \cite{rfc6056}, Random-Increments Port Selection, is likely to be vulnerable to a variant of the attack we described in this paper.


\bibliographystyle{plain} 
\bibliography{DoS,NetSec,rfc,miscellaneous}

%

\end{document}